\documentclass[12pt,a4paper]{article}

\usepackage{placeins}

\usepackage[utf8]{inputenc}
\usepackage[T1,T2A]{fontenc}
\usepackage[russian,english]{babel}

\usepackage{amssymb,amsmath}
\usepackage{slashed}
\usepackage{mathtools}
\usepackage{feynmf}

\usepackage{mathrsfs}
\usepackage{graphicx}
\usepackage{simpler-wick}

\usepackage[export]{adjustbox}
\usepackage[colorinlistoftodos]{todonotes}
\usepackage[colorlinks=true, allcolors=blue]{hyperref}
\usepackage{hyperref}
\usepackage{authblk}

\usepackage{tikz}
\usepackage{subfigure}

\usepackage{indentfirst} 
\usepackage{gensymb}

\graphicspath{{figures/}}

\usepackage{csquotes}
\usepackage[style=nature,
sorting=none,
language=autobib,
autolang=other,
doi=false,
url=false
]{biblatex}
\AtEveryBibitem{\clearfield{month}}
\AtEveryBibitem{\clearfield{issue}}

\usepackage{titlesec}
\titleformat{\section}{\Large\bfseries}{}{0pt}{\Large}{}
\titlespacing{\section}{1.25cm}{0pt}{0pt}
\titleformat{\subsection}{\large\bfseries}{\thesubsection ~~}{0pt}{\large}{}
\titlespacing{\subsection}{1.25cm}{0pt}{0pt}
\titleformat{\subsubsection}{\bfseries}{\thesubsubsection ~~}{0pt}{}{}
\titlespacing{\subsubsection}{1.25cm}{0pt}{0pt}
\renewcommand{\thesection}{}
\renewcommand{\thesubsection}{\arabic{subsection}}

\usepackage{multirow}

\usepackage{aas_macros}

\title{
Reconstruction of the dark matter density profile from cosmic positron anomaly data}
\author[1,2,*]{K.M. Belotsky}
\author[1]{F.V. Kostromin}
\author[1]{M.L. Solovyov}

\affil[1]{National Research Nuclear University MEPhI (Moscow Engineering Physics Institute), 115409, Kashirskoe shosse 31, Moscow, Russia}
\affil[2]{Novosibirsk State University, 630090, Pirogova street 1, Novosibirsk, Russia}
\affil[*]{\small E-mail: k-belotsky@yandex.ru}
\date{}

\begin{document}

\maketitle

\begin{abstract}
In this work we continue our investigations of the possibility of explanation of the positron anomaly (PA) in cosmic rays with the help of annihilating or decaying dark matter (DM) component by varying its space distribution. In the contrast of our previous studies, where we first assumed some specific spatial distribution of DM component and looked at how it agrees with data, here we solve, in some sense, the inverse problem: we search for distribution, in a mathematical way, which satisfies observational data. A unique algorithm has been implemented which, using linear algebra and adaptive grid methods, adjusts distribution to the data. It allows telling in principle whether or not is possible to solve PA problem by variation of spatial distribution of DM sources. A positive result has been formally obtained. A class of solutions can be identified. Though the distributions obtained at the chosen injection spectra may seem slightly realistic, nonetheless it demonstrates a quite powerful possibility in explaining PA that could be realized in more realistic models. 

    
\end{abstract}

\section{Introduction}\label{s:intro}
Cosmic ray (CR) mysteries such as the positron anomaly (PA) \cite{Pamella,PhysRevLett.110.141102,PhysRevLett.113.121101} belong to the most important astrophysical problems, the solution to which can shed light on fundamental issues such as the dark matter (DM) problem. CR spectra may give unique information on non-gravitational properties of DM. Problem of PA has attracted a lot of attention, and different approaches to its solution were elaborated (based on DM \cite{2024arXiv240601705C}, pulsars \cite{HAWC:2017kbo,Philippov:2020jxu,Linares:2020fck, Orusa:2021tts}, local CR overdensity (bubble) \cite{2022Natur.601..334Z}, resent close SN \cite{Kachelriess:2018ser} and maybe other). None can be considered ultimate. 

This work continues our long series of the works \cite{Belotsky:2016tja,Alekseev:2016off, Belotsky:2018vyt,2019PDU....2600333B,2020arXiv201104425S, ICPPA-2022,2025PPN....56..599B} attempting to explain PA with the help of DM annihilation or decay. To solve the problem of PA but simultaneously not to contradict to data on gamma-ray background (IGRB) \cite{Ackermann:2014usa, DiMauro:2016cbj}, we ``play'' with spatial distribution of the supposed small component of DM (producing positrons and gamma). All the previous works were based on assumption of some specific forms of spatial distribution of the small DM component (in form of disk, circles, spirals, and baryon-like arms in Galaxy), and then it was tested by comparison with the observational data. Changing hypotheses on the form of distribution was accounted for by updating data on IGRB (basically by taking into account for blazars contribution), which increased tension with the data, and by the desire to use a more natural distribution. In the last work [to be published in Particles], it has been shown that baryon-like distribution of DM sources is not able to describe data on cosmic positrons and gamma-rays. Here we solve the inverse task. We search for a form of the space distribution from the data. Mathematical algorithm is implemented which is based on linear algebra and adaptive grid methods. This makes it possible to find out an answer to the question, in principle, whether a solution to the given problem can be found in this way.  

In this work, injection (initial) spectra of positions and gamma have been obtained with MC-generator Pyhtia \cite{2006JHEP...05..026S} and also by extrapolating them analytically. Cosmic positron fluxes at the Earth have been obtained with the help of code GLAPROP \cite{GALPROP}, gamma-ray fluxes were calculated manually.

\section{Search for the spatial distribution with algorithm self-adjusting it to the data}\label{s:kostromin}

\subsection{Methodology}

The core concept of the subsequent analysis leverages the additive property inherent to the problem under investigation. Specifically, if 
the Galactic space containing sources is partitioned into several regions, the combined spectrum from the entire Galaxy will equal the sum of spectra generated by each individual region.

Accordingly, 
the following method to construct the target density profile is proposed:
\begin{enumerate}
\item \textbf{Partition of Galactic space} into regions ${D_{1}, D_{2}, \dots, D_{N}}$ (e.g. a rectangular grid, concentric rings, or arbitrary shapes).
\item \textbf{Assign unit density to each region:} For every $D_i$, define a density profile where $\rho_i = 1$ $\text{GeV}/\text{cm}^3$ inside $D_i$ and $0$ elsewhere. Compute the resulting positron and gamma-ray spectra for each profile:
$$
\rho_{i}(x, y, z) = \begin{cases}
1, \;\;\;(x, y, z) \in D_{i} \\ 
0, \;\;\;(x, y, z) \notin D_{i}
\end{cases}
\;\;\;\;\;\;\;\;\;\;i \in[0, N]
$$
\item \textbf{Express the target spectrum} (experimental data) as a linear combination of these spectra with positive coefficients ${k_{1}, k_{2}, \dots, k_{N}}$.
\item \textbf{Reconstruct the density:} In this work, we consider the annihilation processes, and the reaction rate is proportional to the square of the density, and as a consequence, the required density in each region is determined as the root of the corresponding coefficient. Thus, we obtain the desired profile in the form:
$$
\rho(x, y, z) = \sum_{i=1}^{N} \sqrt{k_{i} } \rho_{i}(x, y, z)
$$
\end{enumerate}

\begin{figure}[h!]
    \centering
    \includegraphics[width = 0.65\textwidth]{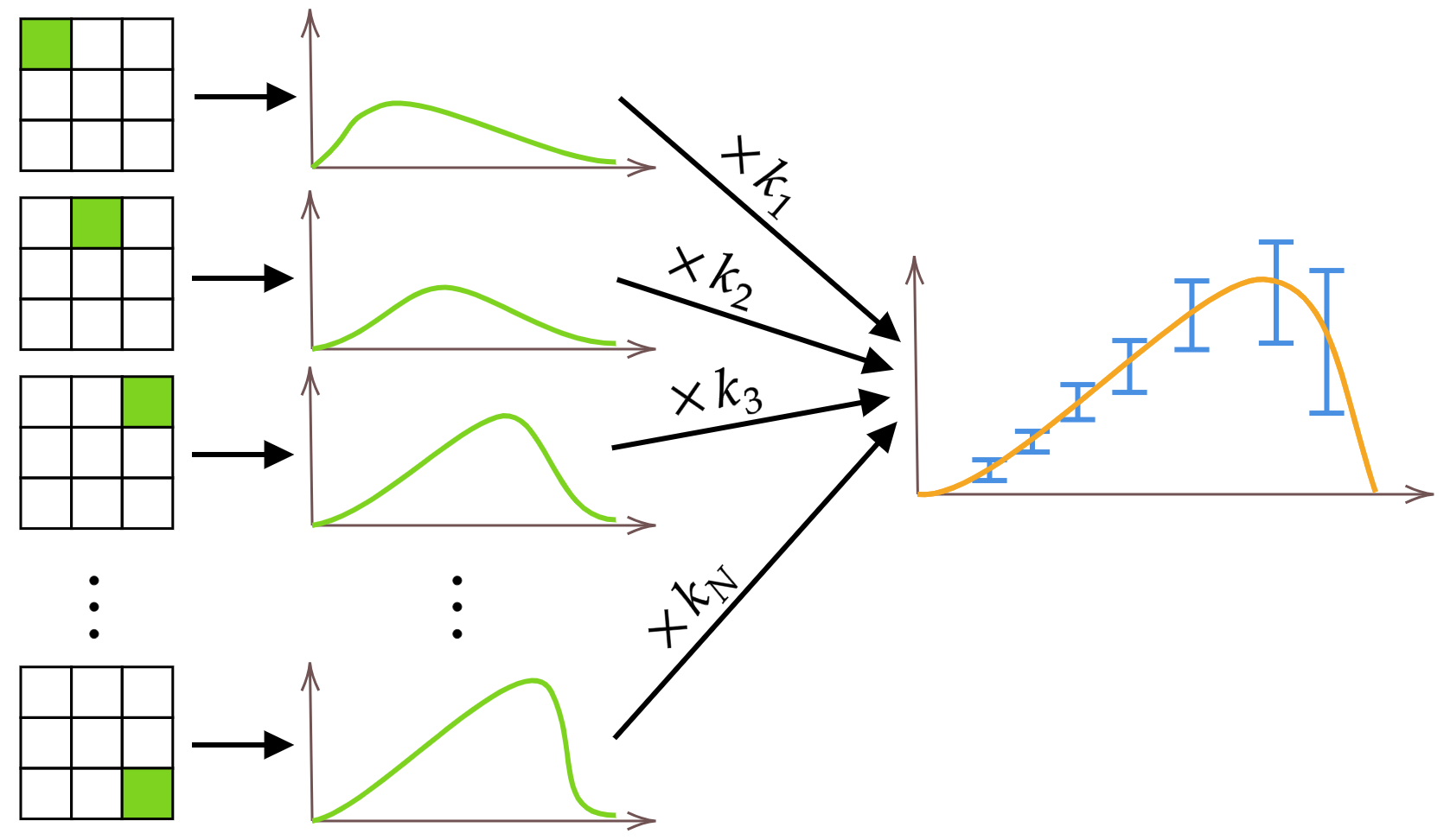}
    \caption{Calculation procedure scheme. On the left, green colour conditionally denotes partitioned areas of unit density, as well as their corresponding spectra in the centre. On the right, orange colour corresponds to the final sum of these spectra compared to experimental data.}  
    \label{fig:decomposition}
\end{figure}

The main idea of the procedure is schematically shown in fig.~\ref{fig:decomposition}.

It should be noted that $\rho_{i}$ profiles do not necessarily have to have the specified form and be localized in a certain region. 
It is possible to choose instead, for example, polynomials, trigonometric functions or another system of functions defined on the entire space as $\rho_{i}$, and the steps outlined above would remain mostly unchanged.

This approach is possible, but it was not used in this paper, since in this case the condition of non-negativity of the density ceases to be equivalent to the condition of non-negativity of the coefficients found in point 3, as a result of which the calculations are significantly complicated. In addition, the approach with areas of unit density seems more visual and understandable from the point of view of analysing the results. 


To calculate the $k_n$ coefficients, the positron and gamma fluxes $\phi_i$ from the obtained spectra corresponding to the energy values of AMS-02 and Fermi-LAT data are combined into following vectors:

$$
D_{i}\;\;\to\;\;\rho_{i}\;\;\to\;\;\begin{bmatrix}
\phi_{i1}^{e^+} \\
\phi_{i2}^{e^+} \\
\vdots \\
\phi_{im}^{e^+} \\
\phi_{i1}^{\gamma} \\
\phi_{i2}^{\gamma} \\
\vdots \\
\phi_{il}^{\gamma}
\end{bmatrix} \equiv \begin{bmatrix}
\phi_{i1} \\
\phi_{i2} \\
\vdots \\
\phi_{iM}
\end{bmatrix}
,
$$

where $i \in [1, N]$, $m$ is the number of experimental points for positrons, $l$ is for gamma-rays, $M=m+l$.

Finding the optimal decomposition of one vector into a set of others with positive coefficients is a linear algebra problem (NonNegative Least Squares problem) that has a known solution \cite{nnls}. Moreover, this solution has already been implemented in software packages such as MATLAB or SciPy (Python).

All these methods minimize the 2-norm of the residual: $||A \vec{k} - \vec{b}|| \to min$, where $A$ is the matrix of vectors along which the decomposition takes place, $\vec{b}$ is the vector that must be obtained, and $\vec{k}$ are the desired coefficients.

In our case, the vector $\vec{b}$ should be composed of experimental points of the positron and gamma spectrum, divided by their errors:

$$
\vec{b}=\begin{bmatrix}
\Phi_{1}^{e^+эксп} / \sigma_{1} \\
\Phi_{2}^{e^+эксп} / \sigma_{2} \\
\vdots \\ 
\Phi_{M-1}^{\gamma\;эксп} / \sigma_{M-1} \\
\Phi_{M}^{\gamma\;эксп} / \sigma_{M}
\end{bmatrix}
$$

The matrix $A$ should be composed of the positron and gamma spectra calculated by us from each region, divided by the corresponding errors of the experimental points:

$$
A = \begin{bmatrix}
\phi_{11}/\sigma_{1} & \phi_{21}/\sigma_{1} & \cdots & \phi_{N1}/\sigma_{1} \\
\phi_{12}/\sigma_{2} & \phi_{22}/\sigma_{2} & \cdots & \phi_{N2}/\sigma_{2} \\
\vdots & \vdots & \ddots & \vdots \\
\phi_{1M}/\sigma_{M} & \phi_{2M}/\sigma_{M} & \cdots & \phi_{NM}/\sigma_{M}
\end{bmatrix}
$$

Then the expression $||A \vec{k} - \vec{b}|| \to min$ will turn into:

$$
\chi^{2} = \sum_{\text{AMS data}} \frac{\Delta \Phi_{e^+}^{2}}{\sigma_{e}^2} + \sum_{\text{IGRB data}} \frac{\Delta \Phi_{\gamma}^{2}}{\sigma_{\gamma}^2}\;\;\to min
,
$$

where $\Delta \Phi_{e^+(\gamma)} = \Phi^{e^+(\gamma)}_{model} - \Phi^{e^+(\gamma)}_{exp} = \sum_{i}k_{i}\phi _i - \Phi^{e^+(\gamma)}_{exp}$.

However, as noted earlier, we do not want to recreate gamma radiation exactly, but only want to not exceed it and therefore should minimize not the expression given above, but:
$$
\chi^{2} =\chi_{\theta}^{2}= \sum_{\text{AMS data}} \frac{\Delta \Phi_{e^+}^{2}}{\sigma_{e}^2} + \sum_{\text{IGRB data}} \frac{\Delta \Phi_{\gamma}^{2}}{\sigma_{\gamma}^2}\textcolor{red}{\theta(\Delta \Phi_{\gamma})}
$$


A simple algorithm was developed, using a standard method for minimizing $\chi^{2}$ in a small number of steps, finds the exact values of the coefficients that minimize $\chi_{\theta}^{2}$. Its operation is described in detail in \ref{ap:norm}.

Thus, we can obtain optimal density values for each region of the selected space partition. This means that the resulting density profile will be more accurate the finer the partition is selected.

As a side note, calculating the spectra from each region in GALPROP requires a significant amount of time, and therefore, simply setting a sufficiently fine partition of the entire space and calculating it is not always possible. Therefore, in some cases, an adaptive grid and step-by-step detailing were used according to the method discussed in \ref{ap:grid}.

The technique described up to this point has several shortcomings. 
First of all, it minimizes the norm that only mimics chi-square, but can not be considered as such from the statistical standpoint. The problem lies in ambiguity of number of degrees of freedom, as the number of regions can be significantly larger than number of AMS datapoints, but the spectres they produce are not really linearly independent, and the sought parameters are constrained. 
Secondly, this algorithm finds only one unique solution, which formally ensures the minimum norm value. However, due to the inevitable error introduced by the finite precision of the calculations, it leads to highly broken and asymmetric density profiles, even if the problem is symmetric.
And while the first issue needs further consideration, the latter could be alleviated by some modifications described below.

To make up for the differences in the spectra of similar density profiles related to accuracy, it was proposed to search not for the density profile corresponding to the minimum chi-square value $\chi^2_{min}$, but for the complete set of density profiles lying in some neighborhood of it $\chi^2_{min}+\Delta \chi^2$.
This can be achieved using an algorithm for sequentially excluding regions with non-zero density, described in detail in \ref{ap:exclusion}.

As a result, we obtain a set of density profiles, each of which satisfies the condition of proximity to $\chi^{2}_{min}$, contains a unique set of regions with non-zero density, and is specified by a set of coefficients $\vec{k}_{i}$. Summing the detected density profiles with certain parameters allows us to construct the final density profile. Accordingly, the final positron or gamma-ray spectrum $\vec{\Phi}$ obtained from the density profile was calculated as the sum of the spectra $\vec{\Phi}_{i}$ from each of the found profiles:

$$
\vec{\Phi}_{i} = 
\begin{bmatrix}
\vec{f_{1}}, \vec{f_{2}}, \vec{f_{3}} \dots\vec{f_{N}}
\end{bmatrix}
\begin{bmatrix}
k_{i1}\\
k_{i2} \\
k_{i3}\\
\vdots \\
k_{iN}
\end{bmatrix}
\text{, and }
\vec{\Phi} = \sum_{i = 1}^{M} \vec{\Phi}_{i} \alpha_{i}
\text{ or }
$$
$$
\vec{\Phi} = 
\begin{bmatrix}
\vec{f_{1}}, \vec{f_{2}}, \vec{f_{3}} \dots\vec{f_{N}}
\end{bmatrix}
 \begin{bmatrix}
k_{11}, k_{21}, k_{31} \dots k_{M 1} \\
k_{12}, k_{22}, k_{32} \dots k_{M 2} \\
k_{13}, k_{23}, k_{33} \dots k_{M 3} \\
\vdots \\
k_{1N}, k_{2N}, k_{3N} \dots k_{M N}
\end{bmatrix}
\begin{bmatrix}
\alpha_{1} \\
\alpha_{2} \\
\alpha_{3} \\
\vdots \\
\alpha_{M}
\end{bmatrix}
,
$$
where $\alpha_{i}>0$ and $\sum_{i}\alpha_{i} = 1$ are arbitrary parameters, $N$ is the number of regions in the chosen partition, $M$ is the number of found sets of coefficients, $\vec{f}_{i}$ are the spectra coming from each of the considered regions at unit density.
Or, in shorter notations, $\vec{\Phi} = FK \vec{\alpha}$.

The choice of the value of $\Delta\chi^2$ was driven by considerations of computational time and the ultimate goal of obtaining density profiles symmetrical about the Sun-Galactic Center axis. Typically, it was on the order of 2–5\% of $\chi^2_{min}$.


Now we have moved from a particular solution to a general solution to the problem, which is parameterized by a set of variables $\alpha_i$. To fix the set of numbers $\alpha_i$ and thus select a single solution, various conditions can be specified, each physically justified in one way or another. In this paper, the condition for the minimum of the highest density value was considered as an example.

\begin{equation}
    \begin{cases}
t = t_{min} \\
K \vec{\alpha} \leq It \\
\sum\alpha_{i} = 1 \\
\alpha_{i}\geq0
\end{cases} ,
\end{equation}
where $t = \rho^2_{max}$ corresponds to the maximum square of the density in the desired profile.This condition, firstly, aims to avoid asymmetric density peaks, and secondly, allows for a unique solution. Other, more complex conditions can also be reduced to solving the problem in this formulation. However, the question of constructing such physically justified conditions remains open at this stage.

There is a huge number of free parameters, so the problem cannot be solved by ordinary fitting. However, in this case of a linear relationship between variables and linear boundary conditions, more specialized methods can be found, 
specifically in the software packages that work with convex linear optimization problems, such as CVXPY (Python).

\subsection{Results}

\subsubsection{Model of an arbitrary flat profile}

At this stage of the work, a thin region lying in the plane of the galaxy with coordinates $x, y \in[-20\text{ kpc}, +20\text{ kpc}]$ and a thickness of 0.6 kpc was considered, divided into $8\times8=64$ identical squares with a side of 5 kpc (Here and below, the partition parameters were largely determined by the desired calculation time). The results are presented for 1000 GeV particles annihilating via the $e^{+}e^{-}$ and $\mu^{+}\mu^{-}$ channels.

\begin{figure}[h!]
    \centering
    \subfigure[]{\includegraphics[width=0.49\textwidth]{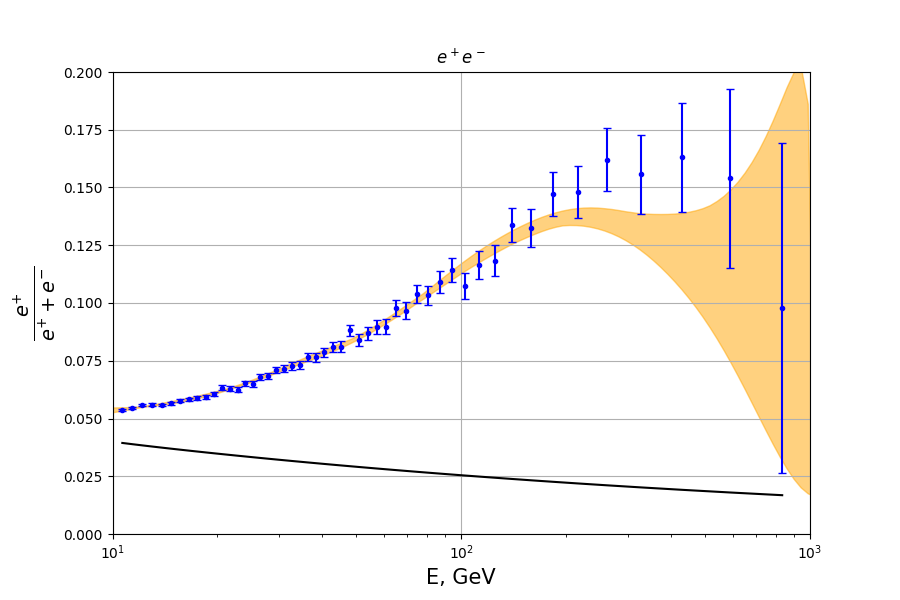}\label{subfig:xy_e_e}}
    \subfigure[]{\includegraphics[width=0.49\textwidth]{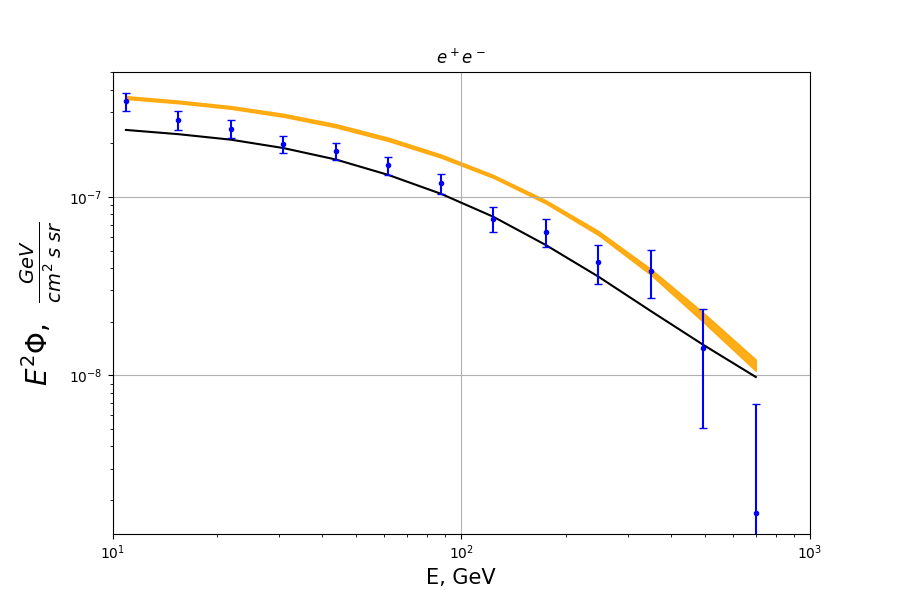}\label{subfig:xy_e_g}}
    
    \subfigure[]{\includegraphics[width=0.49\textwidth]{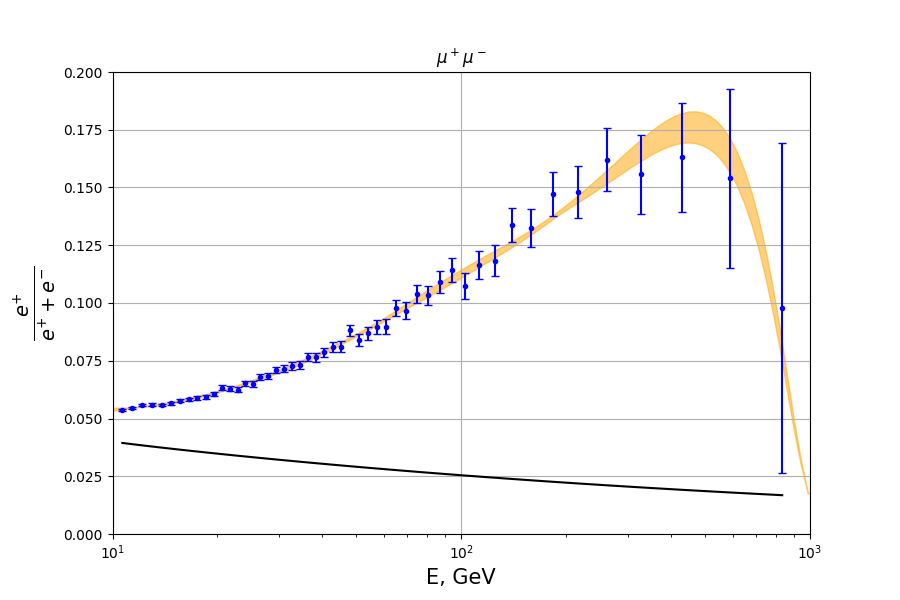}\label{subfig:xy_mu_e}}
    \subfigure[]{\includegraphics[width=0.49\textwidth]{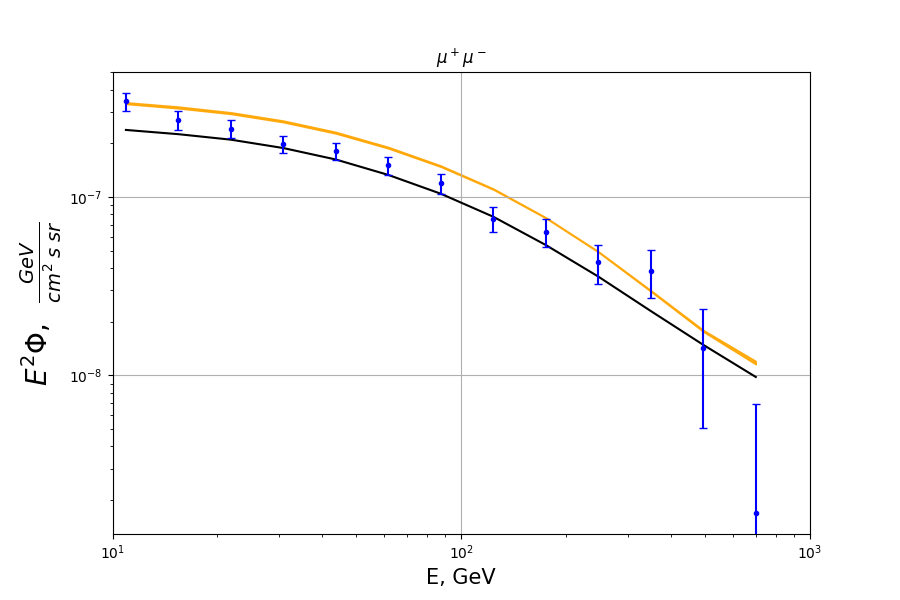}\label{subfig:xy_mu_g}}
    
    \caption{Positron fluxes \subref{subfig:xy_e_e}, \subref{subfig:xy_mu_e} and $\gamma$-radiation \subref{subfig:xy_e_g}, \subref{subfig:xy_mu_g} for all possible values of parameters $\alpha_i$ compared to AMS-02 and Fermi-LAT data, respectively. Cases of annihilation via $e^{+}e^{-}$ channel \subref{subfig:xy_e_e}, \subref{subfig:xy_e_g} and via $\mu^{+}\mu^{-}$ channel \subref{subfig:xy_mu_e}, \subref{subfig:xy_mu_g}.  
    $\Delta \chi^{2}$ is 10\% of $\chi^{2}_{min} = 2.5$ for the $e^+e^-$channel and $\chi^{2}_{min} = 1.3$ for the $\mu^+\mu^-$ channel. Black lines show used background.}
    \label{fig:xy_eg}
\end{figure}

Fig.~\ref{fig:xy_eg} shows the entire set of positron and gamma-ray fluxes that can be obtained for all possible values of the parameters $\alpha_{i}$.

As a first step, it was decided to obtain the upper limit on dark matter density for each region. To do this, it is sufficient to find the maximum coefficient in each row in the matrix $K$. The results are shown in fig.~\ref{fig:rz_up}.

\begin{figure}[h!]
    \centering
    \subfigure[]{\includegraphics[width=0.85\textwidth]{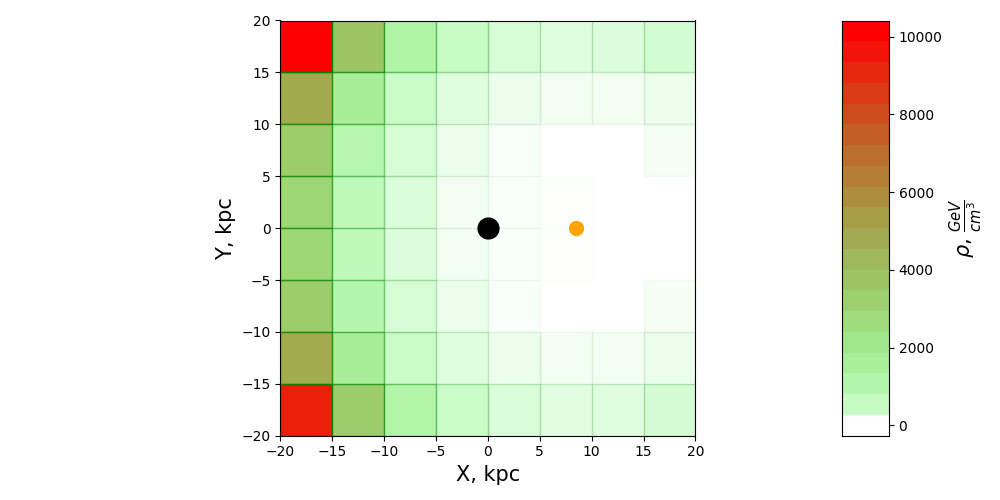}\label{subfig:xy_e_up}}
    \subfigure[]{\includegraphics[width=0.85\textwidth]{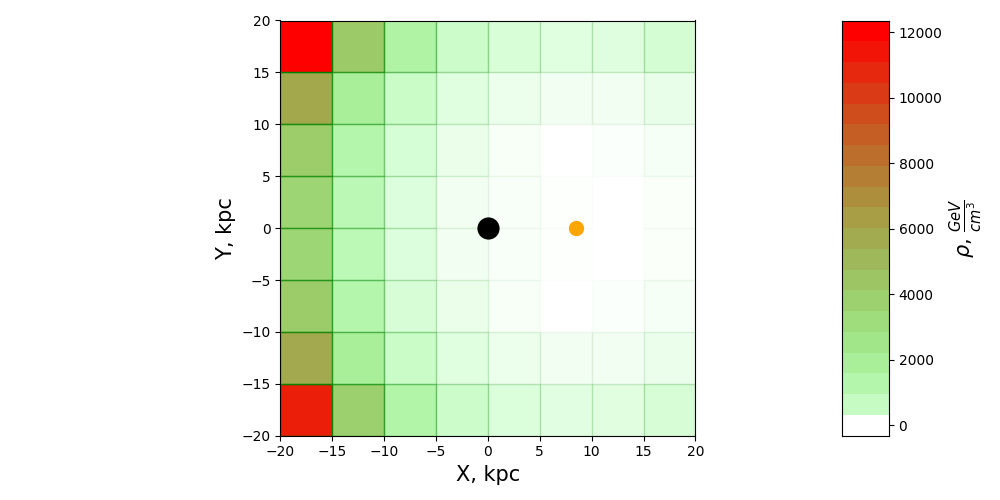}\label{subfig:xy_mu_up}}
    \caption{Upper bound on the source density obtained by choosing the parameter $\Delta \chi^{2}$ equal to 10\% of $\chi^{2}_{min}$. Cases of annihilation via $e^+e^-$ \subref{subfig:xy_e_up} and $\mu^+\mu^-$ \subref{subfig:xy_mu_up} channels.}
    \label{fig:xy_up}
\end{figure}

\begin{figure}[h!]
    \centering
    \subfigure[]{\includegraphics[width=0.85\textwidth]{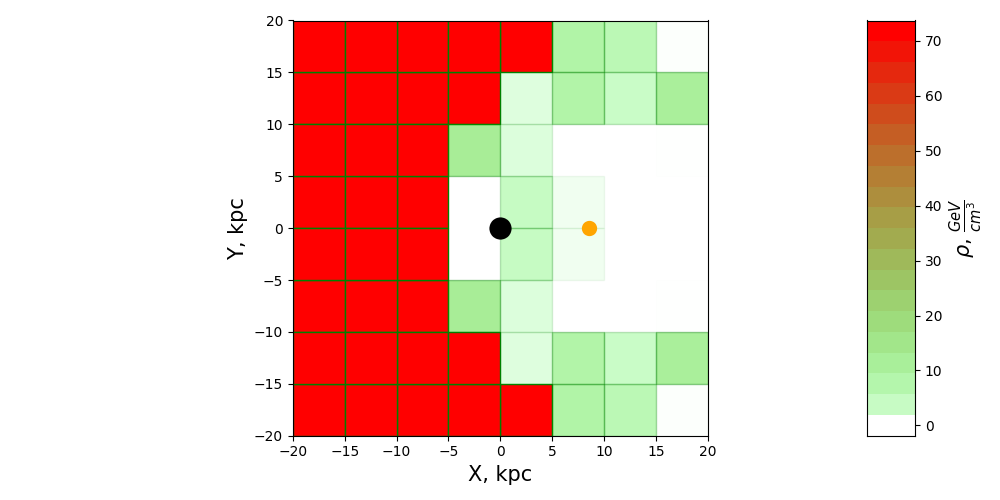}\label{subfig:xy_e}}
    \subfigure[]{\includegraphics[width=0.85\textwidth]{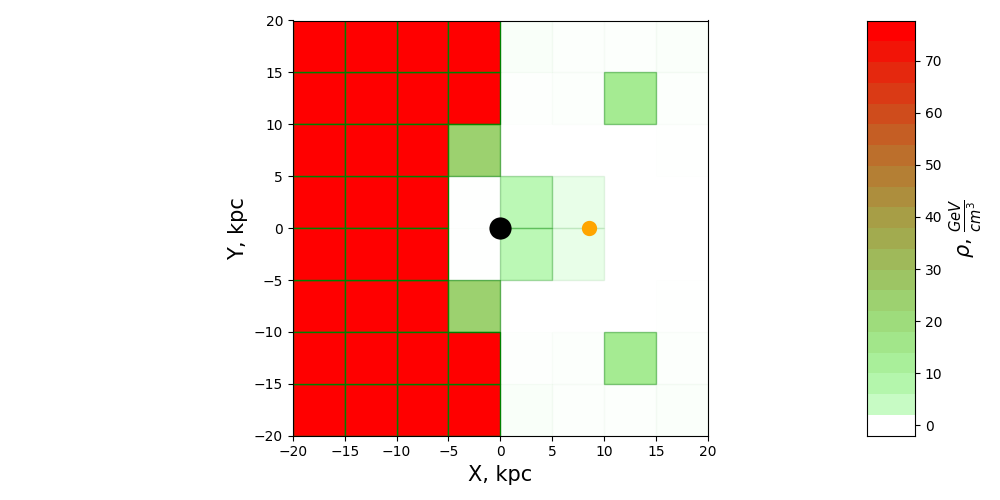}\label{subfig:xy_mu}}
    \caption{The density profile obtained by imposing the condition of the minimum of the highest density. Cases of annihilation via $e^+e^-$ \subref{subfig:xy_e} and $\mu^+\mu^-$ \subref{subfig:xy_mu} channels.}
    \label{fig:xy_single}
\end{figure}

Secondly, the composite density profile was constructed from the set using the conditioned described in previous section.The results are shown in fig.~\ref{fig:xy_single}.

\subsubsection{Model of an arbitrary centrally symmetric profile}

It is clear from the previous section that when considering an arbitrary flat profile within the framework of this method, the density estimates that can be obtained are extremely weak in regions far from the observer. Consideration of a centrally symmetric profile will give us stronger constraints on the source density in regions located further from the Solar System, in particular on the other side of the Galactic Center.

The transition to such a consideration is not difficult, since the methods of this work do not depend on the form of the partition. In this section, 
a region with a radius of $r \in [0 \text{ kpc}, 18 \text{ kpc}]$ around the Galactic Center
and a height of $z \in [-3 \text{ kpc}; 3 \text{ kpc}]$ is considered. The division was made along the $r$ and $z$ axes into $16 \times 8 = 128$ rings with a width of $\Delta r = 1.125\text{ kpc}$ and a height of $\Delta z = 0.75\text{ kpc}$. In this case, to obtain more readable results the adapting grid partitioning was implemented. 

\begin{figure}[h!]
    \centering
    \subfigure[]{\includegraphics[width=0.49\textwidth]{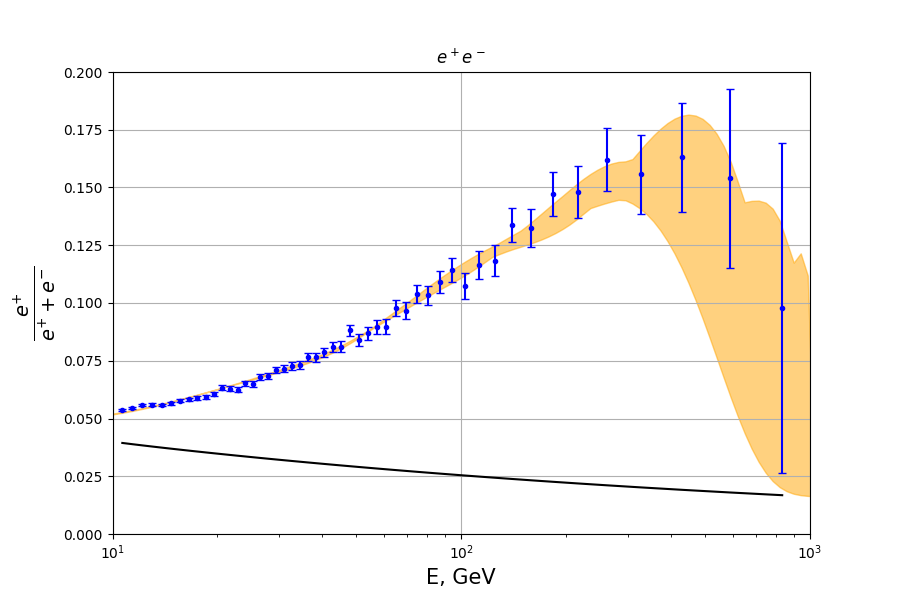}\label{subfig:rz_e_e}}
    \subfigure[]{\includegraphics[width=0.49\textwidth]{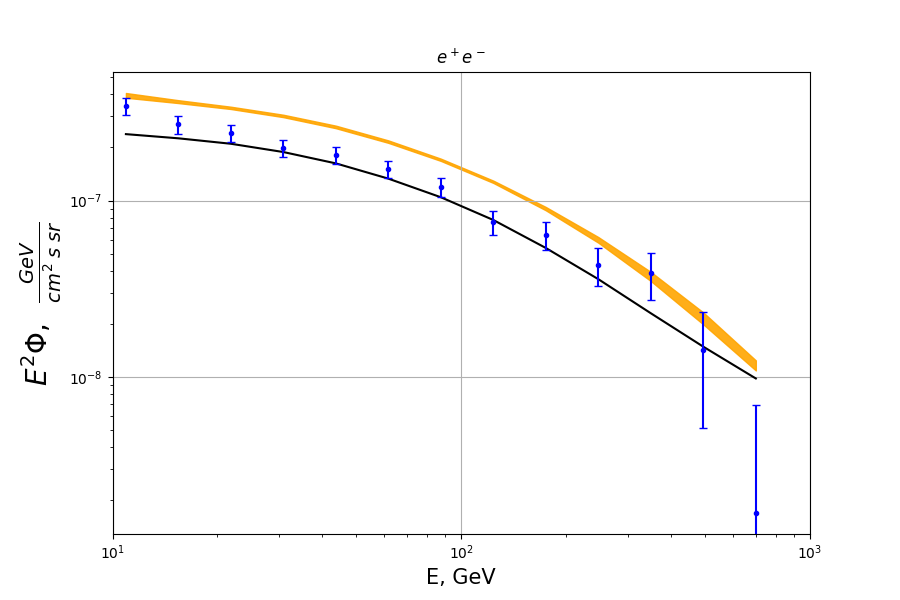}\label{subfig:rz_e_g}}

    \subfigure[]{\includegraphics[width=0.49\textwidth]{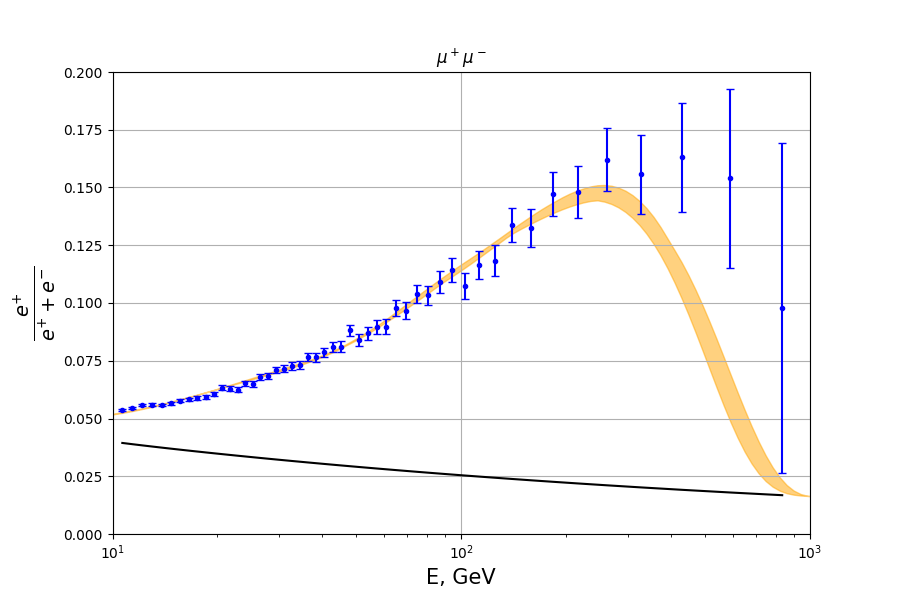}\label{subfig:rz_mu_e}}
    \subfigure[]{\includegraphics[width=0.49\textwidth]{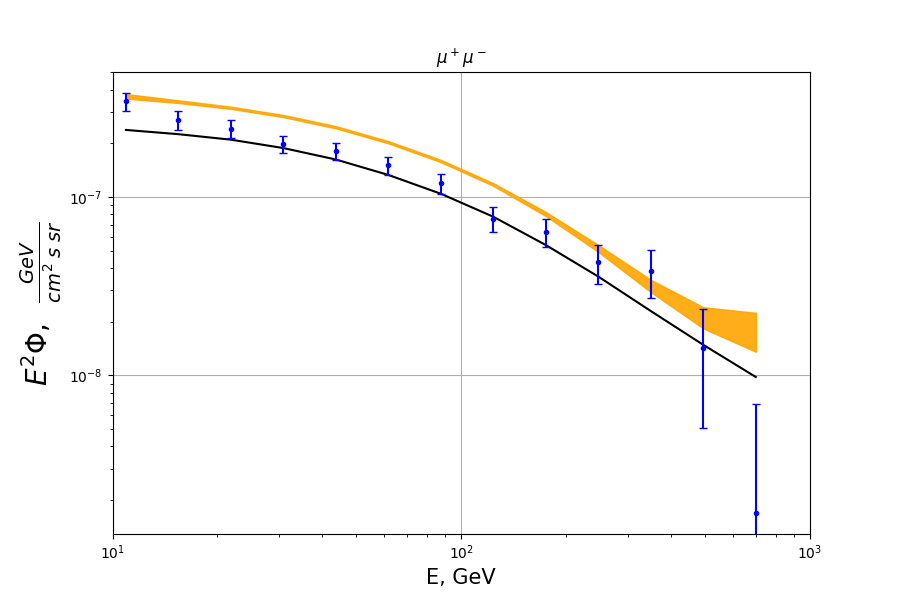}\label{subfig:rz_mu_g}}
    
    \caption{Set of positron fluxes \subref{subfig:rz_e_e}, \subref{subfig:rz_mu_e} and $\gamma$-radiation \subref{subfig:rz_e_g}, \subref{subfig:rz_mu_g} for all possible values of parameters $\alpha_i$ compared to AMS-02 \cite{PhysRevLett.110.141102,PhysRevLett.113.121101,AGUILAR20211} and Fermi-LAT \cite{Ackermann:2014usa} data, respectively. Cases of annihilation via $e^{+}e^{-}$ channel \subref{subfig:rz_e_e}, \subref{subfig:rz_e_g} and via $\mu^{+}\mu^{-}$ channel \subref{subfig:rz_mu_e}, \subref{subfig:rz_mu_g}. 
    $\Delta \chi^{2}$ is 10\% of $\chi^{2}_{min} = 3.3$ for the $e^+e^-$channel and $\chi^{2}_{min} = 3.0$ for the $\mu^+\mu^-$ channel. Black lines show the background used.}
    \label{fig:rz_eg}
\end{figure}

\begin{figure}[h!]
    \centering
    \subfigure[]{\includegraphics[width=0.85\textwidth]{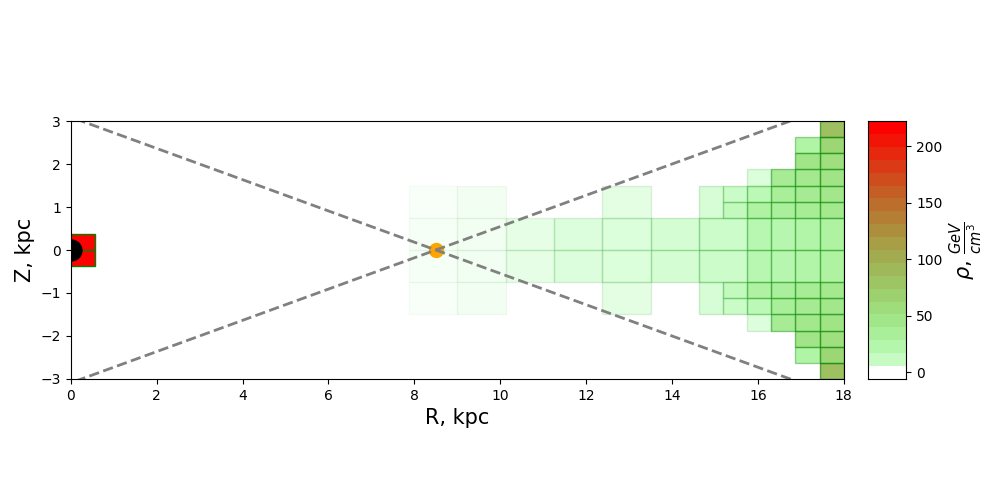}\label{subfig:rz_e}}
    \subfigure[]{\includegraphics[width=0.85\textwidth]{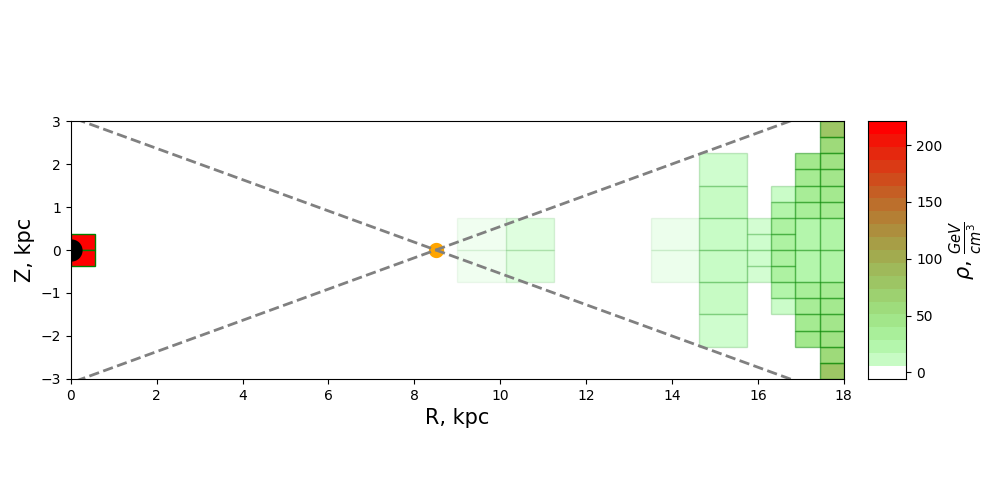}\label{subfig:rz_mu}}
    \caption{Upper bound on the source density obtained by choosing the parameter $\Delta \chi^{2}$ equal to 10\% of $\chi^{2}_{min}$. Cases of annihilation via $e^+e^-$ \subref{subfig:rz_e} and $\mu^+\mu^-$ \subref{subfig:rz_mu} channels. Contrarily to the Fig.\ref{fig:xy_up}, here RZ coordinates are used instead of XY.}
    \label{fig:rz_up}
\end{figure}

Similarly to the previous 
, the sets of obtained spectra of positron and gamma radiation are shown in fig.~\ref{fig:rz_eg}, and the upper limit for the density of sources in fig.~\ref{fig:rz_up}.

\FloatBarrier
\section{Conclusion}\label{s:conclusion}
This work is devoted to the search for solution of the positron anomaly problem in cosmic rays. The search is undertaken by variation of a space distribution of cosmic positron sources. It is assumed that they are accounted for by small component of annihilating (or, with minor corrections, decaying) dark matter forming small structures within disk of the Galaxy. Here we were solving contrarily to the previous works the inverse task. Spatial distribution of the DM component was tried to be derived form the observational data directly. Unique algorithm has been implemented, using linear-algebra and adapting-grid methods. A positive result has been formally obtained. Some class of solutions has been found, see e.g. figures~\ref{fig:xy_up},\ref{fig:xy_single},\ref{fig:rz_up}. Though the distributions obtained at the chosen injection spectra may seem slightly realistic (the sources are concentrated mainly on the periphery of the Galaxy), nonetheless it demonstrates a quite powerful possibility in explaining PA that could be used for some concrete and more realistic models.  



\section*{Acknowledgements}
The work was supported by the Russian Science Foundation grant no.\ 24-22-00325 “Search for an explanation of the positron anomaly in cosmic rays by means of dark matter.”

We would like also to thank our colleagues (A.Kirillov, S.Rubin, V.Nikulin, V.Ganhi, V.Stasenko) for interest to the work and useful discussions.

\printbibliography

\appendix

\renewcommand{\thesection}{Appendix \Alph{section}}
\titleformat{\section}{\Large\bfseries}{\thesection~~}{0pt}{\Large}{}

\section{Norm modifications}
\label{ap:norm}

We have a method that minimizes the deviation norm: $||A \vec{k} - \vec{b}|| \to min$, where $A$ is a matrix of vectors along which the decomposition takes place, $\vec{b}$ is the vector that needs to be obtained, $\vec{k}$ are the desired coefficients. Thus:

$$
\chi^{2} = \sum_{AMS} [(A \vec{k})_{i} - b_{i}]^{2} + \sum_{IGRB} [(A \vec{k})_{i} - b_{i}]^{2}\;\;\to min
$$
Our task is to move on to minimizing the expression:
$$
\chi^{2}_{\theta} = \sum_{AMS} [(A \vec{k})_{i} - b_{i}]^{2} + \sum_{IGRB} [(A \vec{k})_{i} - b_{i}]^{2}\textcolor{red}{\theta((A \vec{k})_{i} - b_{i})}
$$

To move from minimizing $\chi^{2}$ to minimizing $\chi_{\theta}^{2}$, we can do the following:
\begin{enumerate}
    \item We minimize $\chi^{2}$ and obtain the coefficients $\vec{k}$.

    \item Next, we consider the vectors $\vec{b}$ and $A\vec{k}$ and simply discard from consideration the points for which $(\vec{b})_{i}>(A\vec{k})_{i}$.

    \item Then we repeat the previous steps for the new truncated vectors, and repeat the entire cycle until all the calculated points related to gamma radiation used in the minimization lie above the experimental ones.
\end{enumerate}

Thus, in a small number of steps (less than the number of gamma points), we obtain the coefficients that minimize $\chi_{\theta}^{2}$.

\begin{figure}[h!]
    \centering
    \includegraphics[width = 0.85\textwidth]{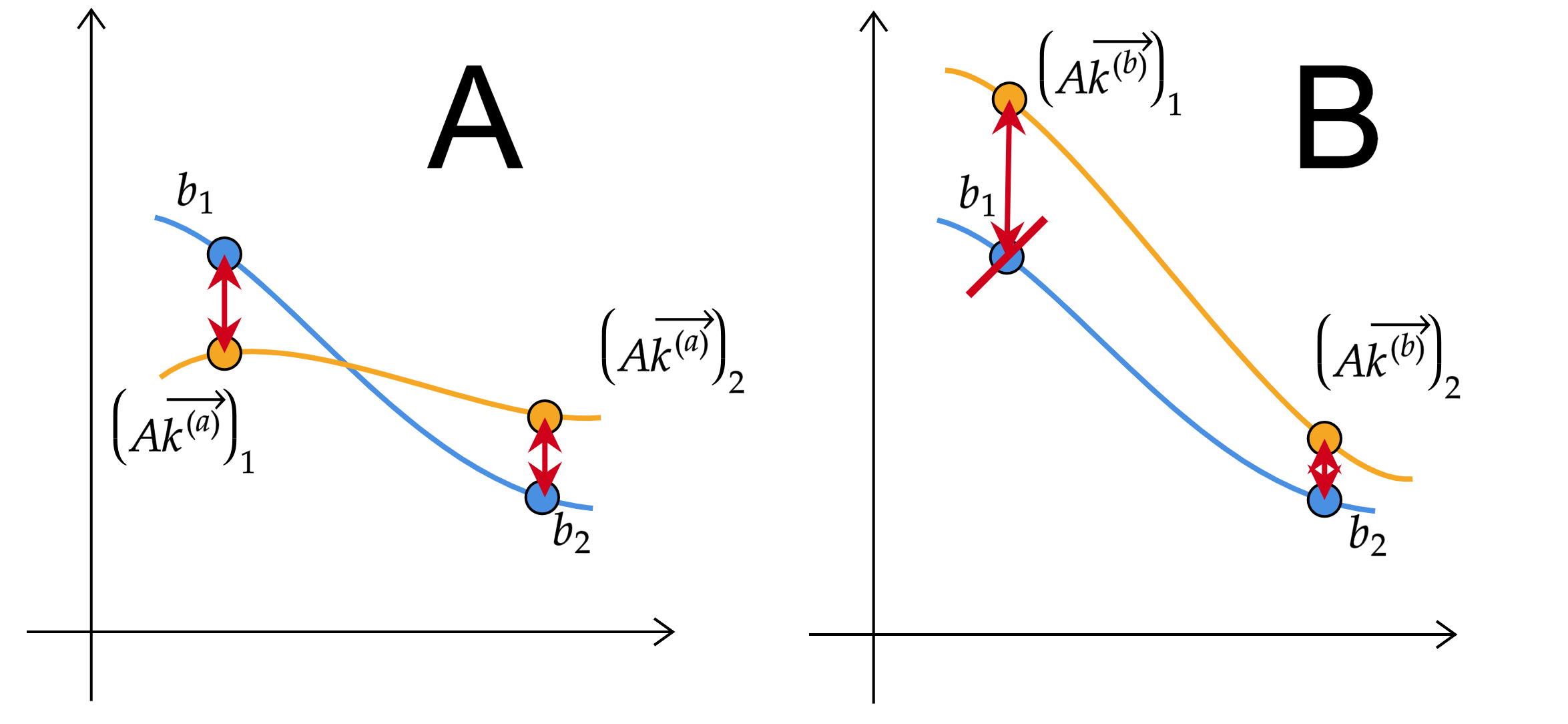}
    \caption{A diagram of the algorithm's applicability proof. Experimental points are shown in blue, and calculated values are shown in orange.}  
    \label{diagram2}
\end{figure}

It may not seem obvious that this procedure leads to the correct result. Let's assume the following situation (using two data points $b_{1}$ and $b_{2}$ as an example): using both points ($b_{1}$ and $b_{2}$), we found the coefficients $\vec{k}^{(a)}$ that minimize $\chi^{2}$, plotted the calculated points $(A \vec{k}^{(a)})_{1}$ and $(A \vec{k}^{(a)})_{2}$, and it turned out that the first experimental point is higher than the calculated one (Figure \ref{diagram2} case A). Then we discarded this point from consideration and found new coefficients $\vec{k}^{(b)}$ for the remaining one ($b_{2}$) that minimize $\chi^{2}$, but it turned out that the point we discarded now lies below the calculated one and should be taken into account (Figure \ref{diagram2} case B). If this is possible, then the method does not work.

It can be shown that this cannot happen (by contradiction).
\begin{itemize}
    \item Let $\Delta_{2}^{(a)} = (A \vec{k}^{(a)})_{2} - b_{2}$ and $\Delta_{2}^{(b)} = (A \vec{k}^{(b)})_{2} - b_{2}$. 
    \item Then, from the fact that only the second point was used in the minimization in case B, it follows that $\Delta_{2}^{(b)} \leq\Delta_{2}^{(a)}$, and in our case, equality is not realized.
    \item Let $\vec{k}(t) = \vec{k}^{(a)}(1-t) + \vec{k}^{(b)}t$ , where t is a parameter varying from 0 to 1.
    \item Then $\Delta_{1} = (A \vec{k}(t))_{1} - b_{1}$ and $\Delta_{2}= (A \vec{k}(t))_{2} - b_{2}$ - depend linearly on t.
    \item $\Delta_{1}^{(a)}<0$ and $\Delta_{1}^{(b)}>0$, then there exists $t_{0} \in (0, 1)$ for which $\Delta_{1}(t_{0})=0$. Moreover, by linearity, $0<\Delta_{2}(t_{0})<\Delta_{2}^{(a)}$.

    \item This means that we have found the coefficients $\vec{k}(t_{0})$ for which $\chi^{2}$ for two points ($b_{1}$ and $b_{2}$) is less than the coefficients $\vec{k}^{(a)}$, which contradicts the initial assumption. Therefore, this cannot happen.
\end{itemize}

This reasoning easily generalizes to a larger number of points and allows us to discard unnecessary points using the method described above.

\section{Exclusion algorithm}
\label{ap:exclusion}
This section is devoted to an algorithm capable of finding all possible density profile variants that are close to
the value of $\chi^{2}_{min}$.

For a clearer presentation of the idea of further work, we will begin with a detailed consideration of a very simplified case. Namely, we will assume that close spectra can only be obtained from regions symmetrical with respect to Solar system - Galactic centre axis.

Let us partition the space under study with a grid and obtain a set of regions. We denote them as $\{a, b, c, d\dots, A, B, C, D \dots\}$, so that the regions $\{a, b, c, d\dots\}$ are symmetric to the regions $\{A, B, C, D\dots\}$.

We already have a fast algorithm that allows us to find one set of coefficients corresponding to the minimum $\chi^{2}$. Suppose we obtained non-zero coefficients for regions $\{a, b, c\}$ and zero for all the others. We denote the resulting density profile as $\rho(\{a, b, c\})$, and denote the densities in each of the regions by $\rho(a), \;\rho(b), \;\rho(c)$, respectively.

Now we exclude the region $a$ from consideration and apply our algorithm for finding optimal densities to the remaining regions. We obtain another variant of interest to us -- $\{A, b, c\}$. Let us thus consider all possible exclusions from the set $\{a, b, c\}$ and obtain 8 profile variants at the output: $\{a, b, c\}$, $\{A, b, c\}$, $\{a, B, c\}$, $\{a, b, C\}$, $\{A, B, c\}$, $\{A, b, C\}$, $\{a, B, C\}$, $\{A, B, C\}$.

Then the general solution is written as:
$$
\rho^{2}(x, y, z) = \alpha_{1}\rho^{2}(\{a, b, c\}) + \alpha_{2}\rho^{2}(\{A, b, c\}) + \dots+\alpha_{8}\rho^{2}(\{A, B, C\})
$$
Where $\alpha_{i}\geq 0$, $\sum_{i}\alpha_{i} = 1$ are arbitrary coefficients. Any chosen set of these values will give us a profile that satisfactorily describes the data.

At this point, a few clarifications should be made.

Firstly, it is the squares of the densities that are added together, i.e. the values of the coefficients found, because, as noted earlier, the process of annihilation of two particles is under consideration. For the decay process, it would be necessary to add the densities to the first power.

Secondly, it is obvious that we could write down the general solution using fewer parameters. For example, in the region $a$ we could take the square of the density $\alpha'_{1} \rho^{2}(a)$ and in the region $A$ take $(1-\alpha'_{1})\rho^{2}(a)$, similarly for pairs of regions $b, B$ and $c, C$. Then we would have only 3 parameters $\alpha'_{1}, \alpha'_{2}, \alpha'_{3} \in[0;1]$. However, with this method we cannot simply add the found profiles as equivalent in terms of spectra, we need to establish equivalence relations between their individual regions or groups of their regions. In the simplified case under consideration, these relations are trivial, but in further work such consideration is not possible.

Thirdly, we can still reduce the number of parameters. After all, it is obvious that, for example:
$$
\rho^{2}(\{A, B, c\}) = \rho^{2}(\{A, b, c\}) + \rho^{2}(\{a, B, c\}) - \rho^{2}(\{a, b, c\})
$$
And then we do not necessarily need to use all 8 profiles. However, with this consideration, the condition of non-negativity of the square of the density does not turn into the requirement of non-negativity of the coefficients $\alpha_{i}$ and the restrictions that we will have to impose on these coefficients, generally speaking, begin to depend on the density in each of the regions. In this simplified case, such consideration is possible, but not feasible in the future.

In this simplified example, we could, of course, not have gone down such a complicated path and easily written out the answer right away. However, we specifically considered the procedure for finding a general solution, which used only the algorithm for finding optimal coefficients and excluding some regions from consideration and did not use knowledge of the symmetry of the problem in any way. Now we can apply it to a more general case.

The point is that by removing, for example, the regions $a, A, b, B$ from consideration and running the algorithm for finding the optimal profile, we can obtain at the output a variant containing, for example, the regions $\{d, e, f, c\}$, which still has an acceptable $\chi^{2}$. It turns out that the sum of the regions $a, b$ included in our initial decomposition can be replaced by some sum of the regions $d, e, f$.

Of the set of spectra from each of the regions that we use, many are "linearly dependent"\ in the sense that the sums of some sets of spectra with positive coefficients can be represented with good accuracy as sums of other sets, again with positive coefficients, and the problem under study has many non-obvious "symmetries".

The algorithm, in this case, should be as following:
\begin{itemize}
\item First, we should consider all possible exclusions from the full set of our regions.

\item Second, for each such exclusion, we should find the best coefficients and the corresponding value of $\chi^{2}$.

\item Next, we need to select all cases in which $\chi^{2}$ value is close enough to the minimum possible one and write out the general solution as their sum, depending on the parameters.
\end{itemize}

It is clear that some of the sets of coefficients obtained in this way will be the same (for example, if we obtained a solution containing only regions $\{a, b, c\}$, then when deleting region $d$, the search for optimal coefficients will not lead to anything new) and many terms in the final sum can be omitted. But it is also clear that by running such an enumeration, we will indeed cover all possible variants of profiles containing different areas and consistent with the data.

However, the number of possible exclusions is $2^{N}$, where $N$ is the number of areas of the order of 100. Obviously, no matter how fast the procedure for finding optimal coefficients is, it is impossible to do it $2^{100}$ times in a reasonable time, and we need to find a way to optimize this search.

\begin{itemize}
\item First, we need to omit from consideration in advance exclusions that obviously lead to the same result.

\item Second, we need to organize the enumeration in the order of increasing $\chi^{2}$, so as not to consider all exclusions, but, moving from the minimum value, stop the enumeration upon reaching a certain boundary.
\end{itemize}

\begin{figure}[h!]
    \centering
    \includegraphics[width = 0.85\textwidth]{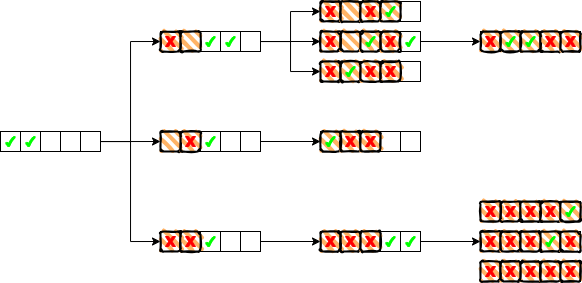}
    \caption{Scheme of enumeration of variants. Red crosses mark the removed regions. Green check marks mark the regions whose coefficients after decomposition turned out to be different from zero. Orange shading marks the areas inaccessible for exclusion during further enumeration.}
    \label{fig:enum}
\end{figure}

An algorithm was developed that satisfies these requirements, and its operation is schematically illustrated in fig.~\ref{fig:enum} using the example of a set of 5 regions.

We start with the best possible decomposition. Next, we consider all possible exclusions of the regions included in it, and all these regions are marked (orange shading in the fig.~\ref{fig:enum}) and become inaccessible for further exclusions. Next, for each exclusion, we again perform the decomposition and find regions with non-zero coefficients. In this case, we are interested only in those of them that were not marked (shaded) earlier. We again consider their possible exclusions and so on, creating a tree of all possible options.

Such an algorithm does not guarantee that identical sets of coefficients will not appear (they must be selected and excluded later). However, it is clear that it does not consider exclusions that obviously lead to the same results (in the given example, it does not consider exclusions containing regions 3, 4, 5 and not containing regions 1 or 2, since they will obviously lead to the original set of coefficients).

Also, such an algorithm does not track how the values of $\chi^{2}$ are related in different branches of the tree. However, it guarantees that, moving further along the tree, we will get $\chi^{2}$ worse at each step than at the previous one. Thus, recursively iterating over such a tree, we can always stop after reaching a certain value of $\chi^{2}_{min} + \Delta \chi^{2}$.

\section{Adapting grid}
\label{ap:grid}

We can construct an approximation of the desired density profile as a piecewise constant function depending on the partition of space. The finer the partition, the more accurate the obtained approximation. However, we cannot always afford to calculate a partition fine enough for the obtained result to be amenable to analysis. Therefore, the idea of using a non-uniform partition and detailing the more "interesting" sections of the profile suggests itself. However, it is impractical to have to start anew with totally different partition if the previous one delivered unsatisfactory results. And so,  it would be beneficial to make the detailing process step-by-step and gradual, so that at any time it would be possible to stop the calculation, see the result and, if necessary, continue further detailing from the last state.



The method used in this work 
is described below for the case of 3D grid consisting of parallelepipeds.


\begin{enumerate}
\item An initial rough uniform partition of space was specified. The observed spectra for each region were calculated using the method described earlier.

\item The optimal density was calculated for each region of the selected partition.
Taking these densities into account, each section was assigned a value proportional to its contribution to the chi-square.

\item Next, the algorithm made a decision on partitioning one of the regions along one of the three axes into two equal daughter regions. The decision at each step was made in such a way as to minimize the difference in values between neighbouring regions and, accordingly, smooth the profile and improve the chi-square.

\item Then, for one of the two obtained daughter regions, the observed spectra were calculated.
For the other, they were found by subtracting the spectra of the calculated daughter region from the spectra of the parent region.

\item In this way, a new, finer division was obtained, and all steps, starting with the second, were repeated again.
\end{enumerate}

\begin{figure}[h!]
    \centering
    \includegraphics[width = 0.85\textwidth]{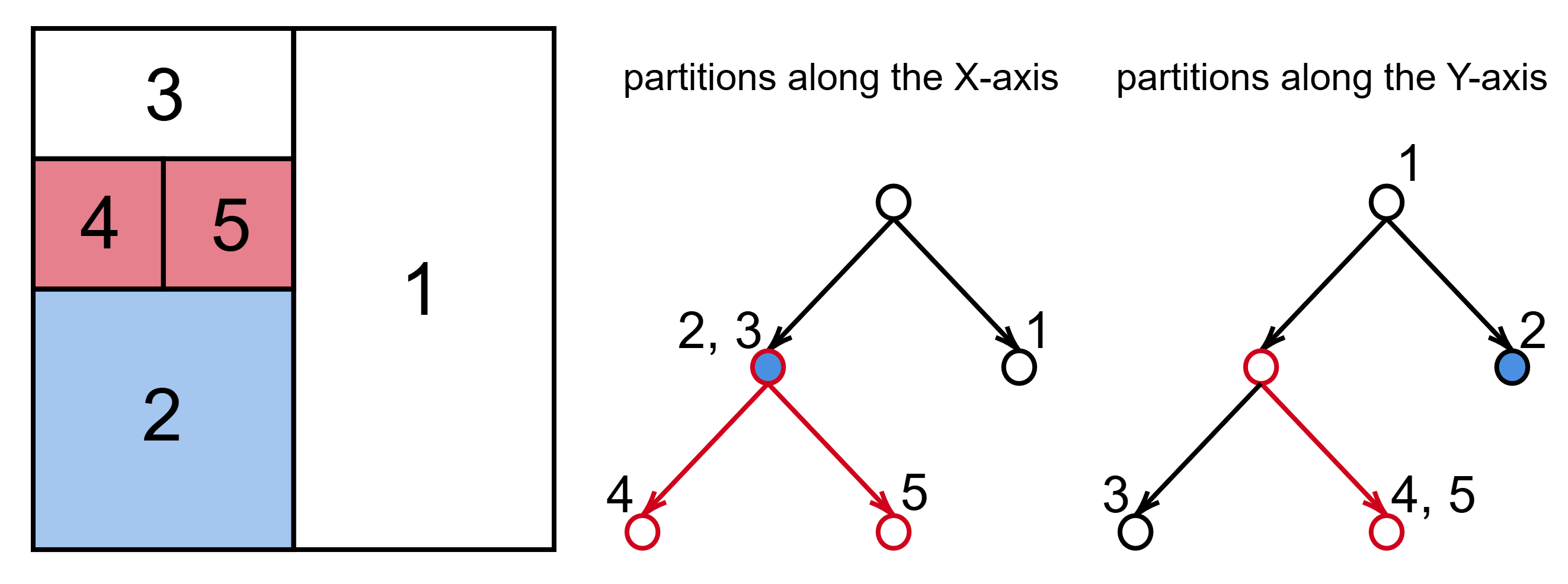}
    \caption{Scheme of work with non-uniform grid.}
    \label{fig:partitions_tree}
\end{figure}

Such a grid was stored as two (for the two-dimensional case) or three (for the three-dimensional case) binary trees of partitions along each of the axes (see fig.~\ref{fig:partitions_tree} for reference).

Each node in each tree is associated with a set of regions, so that the intersection of any sets of any two nodes from different trees contains at most one element. This approach to working with the grid, in addition to the convenience of data storage, allows for each region to quickly select sets of regions bordering it on each side, which minimizes the search time.

Simple selection rules can be formulated. For example, we want to find the set of regions bordering region 2 from above. Then from the partition tree along the X axis, we need to take all nodes that are children of the node containing region 2 (highlighted in color in fig.~\ref{fig:partitions_tree}). We get the set of regions {2, 3, 4, 5}. In the partition tree along the Y axis, from the node containing region 2, we should go up one vertex, go down one step to the left, and then go down the tree, turning only to the right. We get the set of regions {4, 5}. The intersection of the resulting sets will give the desired set {4, 5}.

These selection rules are easily generalized to the three-dimensional case with three trees.

Now, for the selected region, knowing the values in all the regions adjacent to it, we can estimate the difference in values along one of the axes, for example, along the $X$ axis, as follows:
\begin{equation*}
    \Delta_{1x} = |X_{l}-X| + |X_{r}-X|,
\end{equation*}
where $X$ is the value in the region under consideration, $X_{l}$ and $X_{r}$ are the average values over a set of regions adjacent to the left and right, respectively.

\begin{figure}[!h]
    \centering
    \includegraphics[width = 0.65\textwidth]{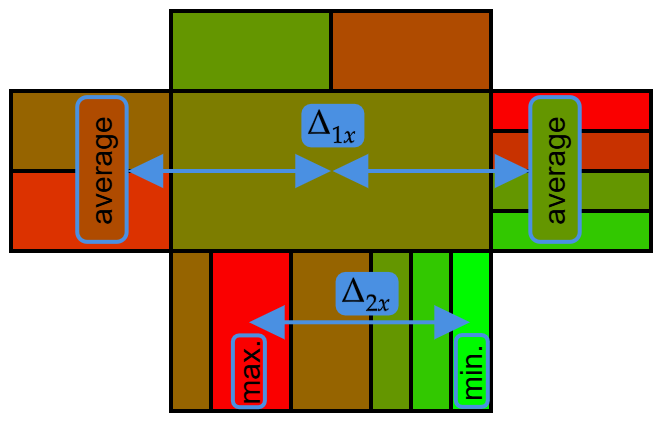}
    \caption{Scheme for determining the values of $\Delta_{1x}$ and $\Delta_{2x}$ for the selected area. Areas that have been assigned a higher value are shown in red, while areas that have been assigned a lower value are shown in green.}
    \label{fig:deltaX}
\end{figure}

The maximum difference in values on the set of regions adjacent to the region on the remaining sides (from the remaining two for the two-dimensional case and from the remaining four for the three-dimensional case) was also considered:
\begin{equation*}
    \Delta_{2x} = X_{t}^{max}-X_{t}^{min},
\end{equation*}
where $X_{t}^{max}$ and $X_{t}^{min}$ are the maximum and minimum values on the specified set of tangent regions (the process of determining the values $\Delta_{1x}$ and $\Delta_{2x}$ is 
outlined in fig.~\ref{fig:deltaX}).

For the region under consideration, the value was determined:\\
\begin{equation*}
    \Delta_{x} = \max[\Delta_{1x}, \Delta_{2x}]*L_{x},
\end{equation*}
where $L_{x}$ is the length of the region under consideration along the $X$ axis.

Thus, for each region, the values $\Delta_{x}, \Delta_{y}, \Delta_{z}$ were calculated. Next, the largest of all these values was selected and the corresponding region was divided along the corresponding axis.

It should also be noted that GALPROP uses a spatial grid with a certain step when numerically modeling particle flows. In this regard, in order to avoid significant errors in modelling, it is necessary to introduce restrictions on partitions by setting the minimum permissible length of regions along each of the axes.


\end{document}